\documentclass[prl,showpacs,superscriptaddress,twocolumn]{revtex4} 
\pdfoutput=1
\usepackage{amssymb,amsmath,graphicx}
\begin{document}
\title{Cascaded exciton emission of an individual strain-induced quantum dot}

\author{F. J. R. Sch\"{u}lein}
\affiliation{Lehrstuhl f\"{u}r Experimentalphysik 1, Universit\"{a}t Augsburg, Universit\"{a}tsstr. 1, 86159 Augsburg, Germany} 
\author{A. Laucht}
\affiliation{Walter Schottky Institut, Technische Universit\"{a}t M\"{u}nchen, Am Coulombwall 3, 85748 Garching, Germany}
\author{J. Riikonen}
\affiliation{Department of Micro and Nanosciences, Micronova, Helsinki University of Technology, PO Box 3500, FIN-02015 TKK, Finland}
\author{M. Mattila}
\affiliation{Department of Micro and Nanosciences, Micronova, Helsinki University of Technology, PO Box 3500, FIN-02015 TKK, Finland}
\author{M. Sopanen}
\affiliation{Department of Micro and Nanosciences, Micronova, Helsinki University of Technology, PO Box 3500, FIN-02015 TKK, Finland}
\author{H. Lipsanen}
\affiliation{Department of Micro and Nanosciences, Micronova, Helsinki University of Technology, PO Box 3500, FIN-02015 TKK, Finland}
\author{J. J. Finley}
\affiliation{Walter Schottky Institut, Technische Universit\"{a}t M\"{u}nchen, Am Coulombwall 3, 85748 Garching, Germany}
\author{A. Wixforth}
\affiliation{Lehrstuhl f\"{u}r Experimentalphysik 1, Universit\"{a}t Augsburg, Universit\"{a}tsstr. 1, 86159 Augsburg, Germany} 
\author{H. J. Krenner}
\affiliation{Lehrstuhl f\"{u}r Experimentalphysik 1, Universit\"{a}t Augsburg, Universit\"{a}tsstr. 1, 86159 Augsburg, Germany}


\begin{abstract}
	Single strain-induced quantum dots are isolated for optical experiments by selective removal of the inducing InP islands from the sample surface. Unpolarized emission of single, bi- and triexciton transitions are identified by power-dependent photoluminescence spectroscopy. Employing time-resolved experiments performed at different excitation powers we find a pronounced shift of the rise and decay times of these different transitions as expected from cascaded emission. Good agreement is found for a rate equation model for a three step cascade.
\end{abstract}

\maketitle

Since the first demonstration of single photon emission from a single quantum dot (QD) \cite{Michler:00} remarkable progress has been made towards direct device applications with high operation speeds and efficiency \cite{Shields:07,Strauf:07}. Moreover, for more advanced concepts required for inherently secure quantum cryptography protocols sources of entangled photon pairs have been demonstrated in this solid-state platform using the emission cascade from the biexciton $2\mathrm X=2e+2h$ to the empty QD via the intermediate exciton $1\mathrm X=1e+1h$ level \cite{Benson:00}. However, entanglement of the polarization of the two emitted photons is lifted due to the anisotropic exchange splitting which separates the intermediate exciton levels into linear polarized doublets. Therefore, post selection of these two decay channels becomes possible and the entanglement is lifted \cite{Akopian:06}. In this letter, we present a series of photoluminescence (PL) experiments performed on a single strain-induced quantum dot (SI-QD). Here, we identify both the exciton and biexciton emission lines which show no evidence of polarization anisotropy or finestructure. Moreover, in time-resolved experiments we observe a shift of the onset of the exciton emission with respect to the biexciton emission for higher pump powers i.e. increasing probability of biexciton generation. This behavior is indicative for a cascaded emission process and can be reproduced by a rate equation model.

\begin{figure}[htbp]
	\begin{center}
		\includegraphics[width=0.8\columnwidth]{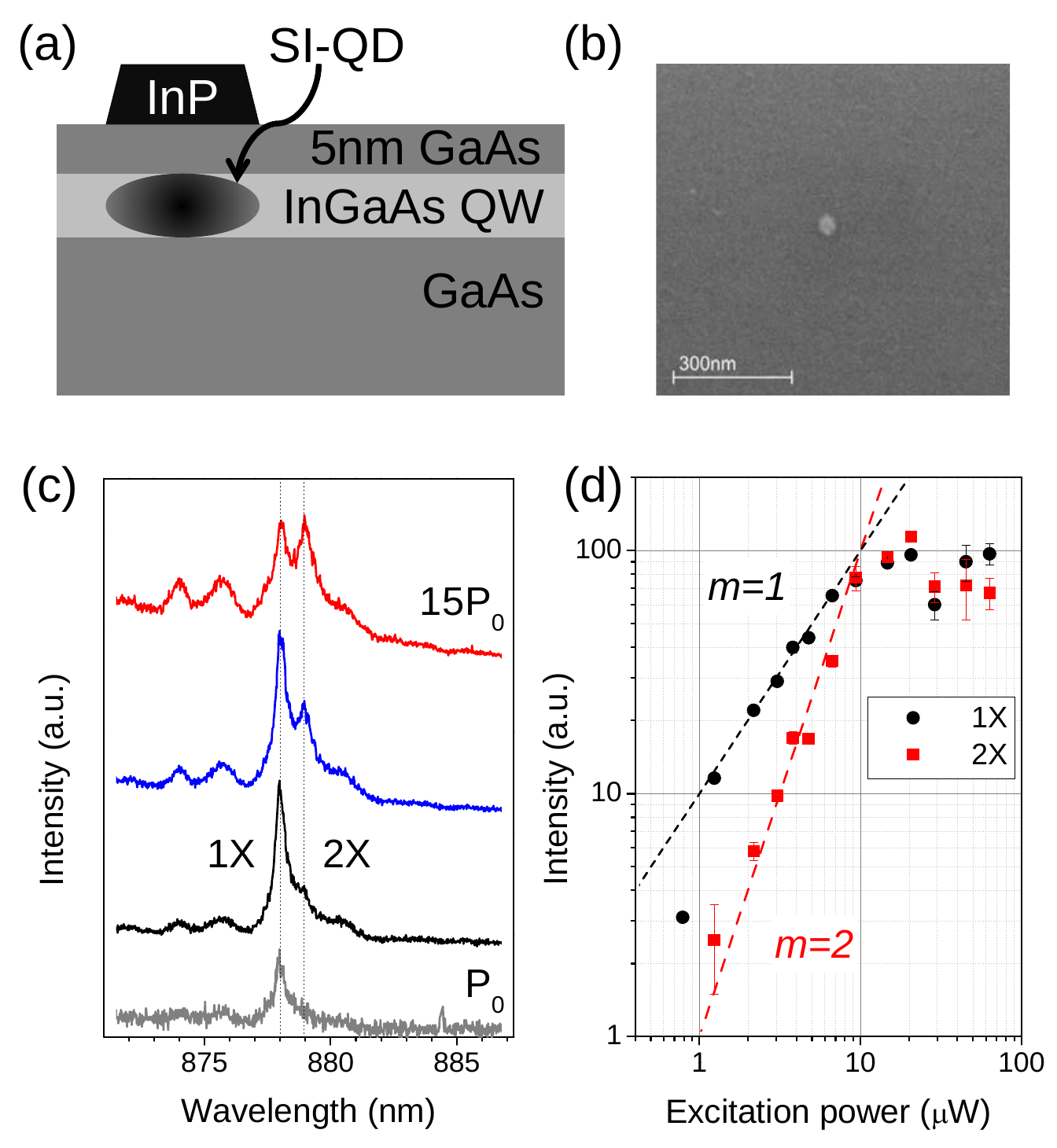}
		\caption{(Color online) (a) Schematic of a SI-QD and (b) SEM image of single SI-QD isolated by selective removal. (c) Power-dependent PL and (d) plot of peak maximum of a single SI-QD to identify the 1X and 2X emission lines.}
		\label{fig:1}
	\end{center}
\end{figure}
The studied sample consists of SI-QDs which are formed by the strain field induced by InP islands in a $4\mathrm{~nm}$ wide In$_{0.1}$Ga$_{0.9}$As/GaAs quantum well (QW) located $5\mathrm{~nm}$ below the sample surface as shown schematically in Fig. \ref{fig:1} (a) \cite{Lipsanen:95}. The ensemble of $60-80\mathrm{~nm}$ diameter InP stressors initially has an areal density of $\sim3\cdot 10^9~\mathrm{cm}^{-2}$ from which individual QDs are isolated by selective removal of the surrounding islands using wet chemical etching by HCl and a mask defined by electron beam lithography. Using this technique we are able to perform spectroscopy on individual SI-QD without using optical near-field or masking techniques \cite{Obermuller:99,Bracker:01}. A typical example of an SEM image of isolated SI-QD is shown in Fig. \ref{fig:1}(b) which can be studied by standard low-temperature $(T=10~\mathrm{K})$ micro-PL. In the experiments presented here, carriers were photogenerated either by a cw-diode laser emitting at $675\mathrm{~nm}$ or a tuneable, pulsed Ti:Sapphire laser (repetition rate $80\mathrm{~MHz}$, pulse duration $2\mathrm{~ps}$). The emission of the QDs was dispersed by a grating monochromator and detected using a liquid nitrogen cooled Si-CCD or a single photon Si-avalanche photodiode detector. For time-resolved experiments a time-to-amplitude converter was used to correlate the excitation pulse of the laser with the output of the single photon detector providing a temporal resolution of $0.35\mathrm{~ns}$.\\

In order to identify the different excitonic transitions of the QD, time-integrated PL spectra were recorded as a function of the excitation laser power, an example being shown in Fig. \ref{fig:1}(c). At low excitation power ($P_0\sim 0.79~\mu\mathrm W)$, a single emission line corresponding to the single exciton (1X) is detected at $878~\mathrm{nm}$. When the excitation power is raised a second sharp line appears at $879~\mathrm{nm}$ i.e. at lower photon energy which we attribute to the biexciton (2X) emission with a binding energy of $1.5\mathrm{~meV}$. At the highest excitation power $(15P_0)$ both lines show approximately equal intensity and additional broader features appear at shorter wavelength arising from transitions in the QD $p$-shell. To further support the assignments of the 1X and 2X lines we analyzed their intensities over a wide range of excitation powers which is plotted in a double-logarithmic representation in Fig. \ref{fig:1} (d). Clearly, the 1X (circles) and 2X (squares) lines show the characteristic linear and quadratic dependence as observed for other QD systems \cite{Brunner:94} and GaSb based SI-QDs \cite{Bracker:01}. In addition, the observed linewidths of 0.8 meV and 1.1 meV for the 1X and 2X emission lines, respectively, are not limited by the spectral resolution of our measurement setup and both lines show no polarization dependence as expected from the isotropic confinement potential \cite{Boxberg:07}. These properties would require no postselection or tuning of the finestructure for the generation of pairs of polarization entangled photons if this emission takes place in a cascade. 

\begin{figure}[htbp]
	\begin{center}
		\includegraphics[width=0.9\columnwidth]{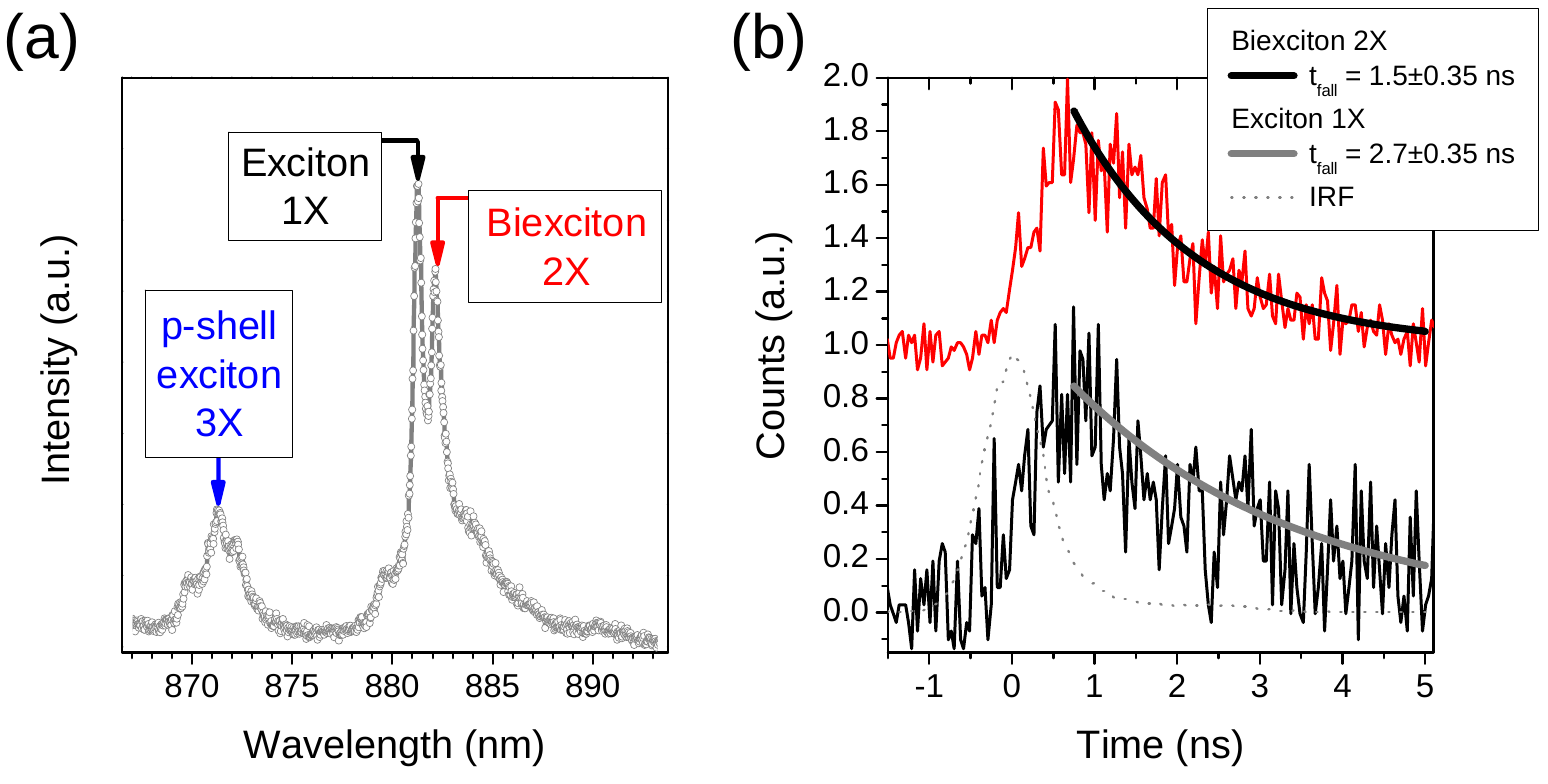}
		\caption{(Color online) (a) Single SI-QD spectrum under pulsed excitation at medium excitation power with the transitions studied in TR-PL marked. (b) TR-PL of the 1X (lower trace) and 2X (upper trace) transitions.}
		\label{fig:2}
	\end{center}
\end{figure}
In order to obtain evidence for a cascaded emission from the 2X via 1X level into the empty QD ground state we performed a time-resolved PL (TR-PL) experiment on the same SI-QD. The wavelength of the Ti:Sa laser was tuned to $818\mathrm{~nm}$ in resonance with the free exciton absorption in GaAs. This leads to better resolved 1X and 2X emission peaks as can be seen in the spectrum shown in Fig. \ref{fig:2} (a). This spectrum was taken at a moderate excitation power of $7~\mu\mathrm{W}$ where the 1X, 2X and triexciton emission in the p-shell are observed as marked in the figure. In a first step we reduced the excitation laser power to $2~\mu\mathrm{W}$ where the 1X emission dominates and only weak 2X emission is detected and measured the PL-transients for these two lines which are presented in Fig. \ref{fig:2} (b). Under these pumping conditions the QD is populated predominantly by a single $e$-$h$ pair and, therefore, we are able to determine the intrinsic PL-decay times of the 1X (upper trace) and 2X (lower trace) states to $1.5\pm0.35\mathrm{~ns}$ and $2.7\pm0.35\mathrm{~ns}$ (bold lines), respectively. The time of the optical excitation can be determined by the maximum of the instrument response function (IRF) which is shown by the dashed gray line. A comparison of these lifetimes and the observed linewidth indicates additional broadening mechanisms present in our system. We cannot exclude that this broadening is superimposed on a finite finestructure splitting, however, no significant change of the emission lineshapes are observed for different polarizations. Furthermore, the nature of these mechanisms determine the coherence and degree of polarization entanglement of the emitted photon pairs.  

\begin{figure}[htbp]
	\begin{center}
		\includegraphics[width=0.75\columnwidth]{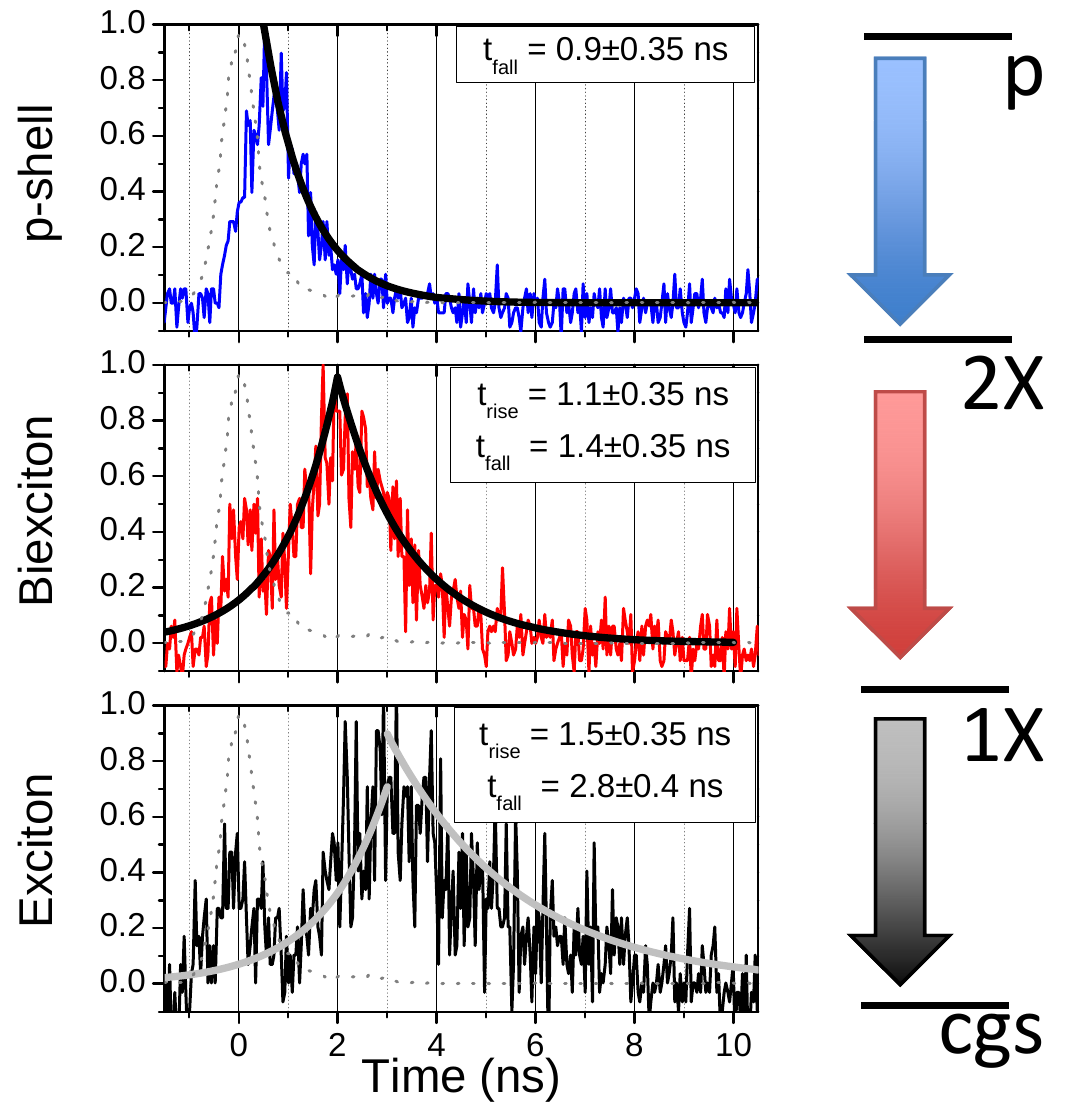}
		\caption{(Color online) Time-resolved PL of the exciton, biexction and triexciton transition (bottom to top panel). A clear shift of the onsets is observed from the p-shell to the 1X emission due to cascaded single photon emission.}
		\label{fig:3}
	\end{center}
\end{figure}
When the excitation power is increased, the biexciton and triexciton generation gets favored over the single exciton as it is the case for the spectrum in Fig. \ref{fig:2} (a). We recorded the PL transients for the 1X, 2X and 3X transitions under these conditions which are shown in  Fig. \ref{fig:3}. When comparing these three traces one clearly notices that only for the 3X decay the maximum number of events is detected shortly after the excitation pulse. In contrast, this maximum is shifted by $2.0\pm0.5 \mathrm{~ns}$ for the 2X and even more by $2.8\pm0.7 \mathrm{~ns}$ for the 1X decay. A detailed analysis of the rise and fall times (shown by bold lines in Fig. \ref{fig:3}) of the three transients show that the fall time of the 3X of $t_{fall,\mathrm{3X}} = 0.9\pm0.35\mathrm{~ns}$ agrees well with the rise time of the 2X transition $t_{rise,\mathrm{2X}} = 1.1\pm0.35\mathrm{~ns}$. The same behavior is observed for the 2X and 1X with  $t_{fall,\mathrm{2X}} = 1.4\pm0.35\mathrm{~ns}$ and $t_{rise,\mathrm{1X}} = 1.5\pm0.35\mathrm{~ns}$. This observation of correlated rise and fall times is a strong indication for a single photon cascade from the 3X level via the 2X to the 1X into the crystal ground state (cgs) of the QD as shown schematically in Fig. \ref{fig:3}.

\begin{figure}[htbp]
	\begin{center}
		\includegraphics[width=0.9\columnwidth]{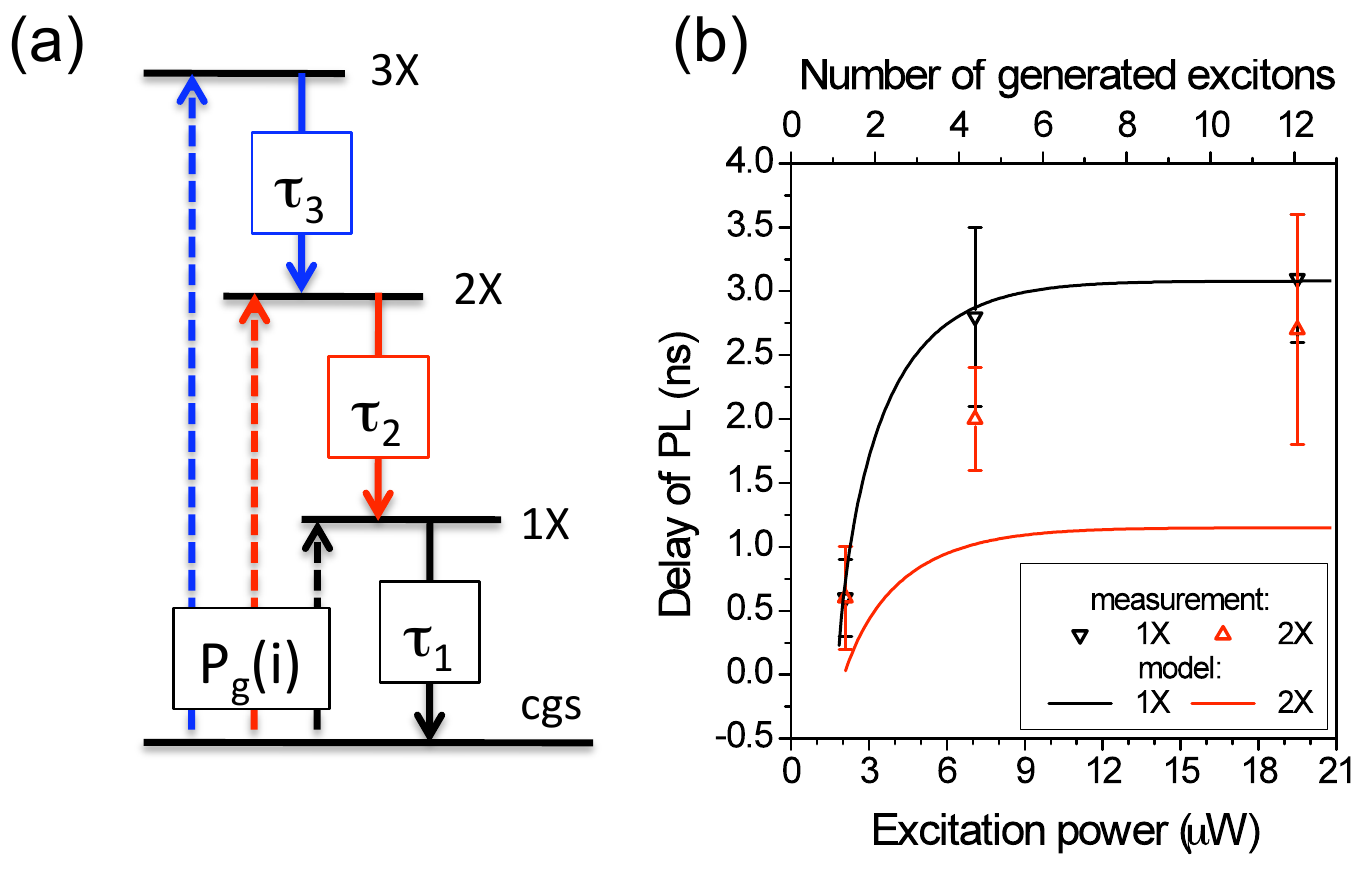}
		\caption{(Color online) (a) Schematic illustration of the terms used in equation \eqref{eq:2}: a level can be populated by direct excitation (dashed arrows) or by the decay of the next upper level (solid arrows). (b) Comparison between the measured delays of the 1X and 2X (symbols) compared to results of a rate equation model (lines).}
		\label{fig:4}
	\end{center}
\end{figure}

To model these results, we applied rate equations for the three levels $i\mathrm{X}$ forming the cascade. The occupation probabilities $n_i$ of level  $i=1,2,3$ is given by
\begin{equation}
	\frac{\mathrm d}{\mathrm dt}n_i=G_i-\frac{n_i}{\tau_i}
	\label{eq:1}
\end{equation}
by the corresponding generation $G_i$ and decay rates $\tau_i^{-1}$. These are shown schematically along with the underlying level structure in Fig. \ref{fig:4} (a).

To solve this set of equations we assumed (i) Poissonian capture of $e$-$h$ pairs independent of the QD occupation (ii) quick carrier relaxation into the QD states and complete transition to the cgs before the next optical excitation pulse \cite{Dekel:00}. Therefore, the generation rate $G_i$ can be written as a sum of the Poisson distribution function $P_g(i)$ ($g$: number of excitons generated in the host matrix by the optical pump) and the decay rate of the next upper level of the cascade
\begin{equation}
G_i=\frac{n_{i+1}}{\tau_{i+1}}+e^{-g}\frac{g^i}{i!}\;\;.
\label{eq:2}
\end{equation}

Using equations \eqref{eq:1} and \eqref{eq:2} and the experimentally determined decay times we are able to calculate the temporal dependence of the PL signal as a function of the number of generated excitons. The points in time of the maximum of these transients at constant $g$ correspond to the PL delay at a given excitation power. These values are plotted in Fig. \ref{fig:4} (b) as solid lines for the 1X (black) and 2X (red/gray) transitions. These results reproduce the experimentally obtained delay times [symbols in Fig. \ref{fig:4} (b)] by assuming a linear dependence between the excitation power and  $g$. The applied conversion is also in agreement with the observed and simulated saturation behavior. In particular good agreement is obtained for the 1X transition and the deviation for 2X at high powers can be related to our limitation to three levels in the model. In this power range we observe also a $\sim 300$ ps delay of the 3X transition indicating that additional higher levels contribute to the cascade.

In summary, we have identified single exciton and biexciton emission of a single SI-QD which are predominantly unpolarized and without fine-structure splitting. TR-PL revealed a delayed onset of lower occupancy states' signal in good agreement with coupled rate equation model thus providing evidence without direct measurement of the cross-correlation function \cite{Moreau:01}. These findings show that SI-QDs might provide a platform for a QD-based source of pairs of polarization entangled photons which does not require elaborate postselection \cite{Akopian:06} or finestructure tuning \cite{Seguin:06,Stevenson:06,Vogel:07} schemes.

This work was financially supported by Deutsche Forschungsgemeinschaft via the Cluster of Excellence Nanosystems Initiative Munich (NIM).

\end{document}